\documentclass[conference]{IEEEtran}

\pdfoutput=1

\IEEEoverridecommandlockouts
\usepackage{cite}
\usepackage{amsmath,amssymb,amsfonts}
\usepackage{algorithmic}
\usepackage{graphicx}
\usepackage{textcomp}
\usepackage{xcolor}
\usepackage{colortbl}
\usepackage{tikz}

\usepackage{amsmath,amssymb,amsfonts}
\usepackage{algorithmic}
\usepackage{graphicx}
\usepackage{textcomp}
\usepackage{xcolor}
\usepackage{longtable}
\usepackage{float}
\usepackage{graphicx}
\usepackage{multirow}
\usepackage{balance}
\usepackage{flushend}
\usepackage{hyperref}
\usepackage{tabularx} 
\usepackage{booktabs} 
\usepackage{hhline}
\usepackage{enumitem}
\usepackage{array}
\usepackage{lipsum}
\newcolumntype{L}[1]{>{\raggedright\let\newline\\\arraybackslash\hspace{0pt}}m{#1}}
\newcolumntype{C}[1]{>{\centering\let\newline\\\arraybackslash\hspace{0pt}}m{#1}}
\newcolumntype{R}[1]{>{\raggedleft\let\newline\\\arraybackslash\hspace{0pt}}m{#1}}
\usepackage{subfigure}

\newenvironment{boxedtext}
    {
    
    \begin{center}

    \begin{tabular}{|p{0.96\linewidth}|}
    \hline
    }
    { 
    \\ \hline
    \end{tabular} 
    
    \end{center}
       }

\def\BibTeX{{\rm B\kern-.05em{\sc i\kern-.025em b}\kern-.08em
    T\kern-.1667em\lower.7ex\hbox{E}\kern-.125emX}}
\begin{document}

\title{Are Code Review Processes Influenced by the Genders of the Participants?}

\author{\IEEEauthorblockN{Sayma Sultana, Amiangshu Bosu}
\IEEEauthorblockA{\textit{Department of Computer Science} \\
\textit{Wayne State University}\\
Detroit, Michigan, USA \\
\{sayma, amiangshu.bosu\}@wayne.edu }
}

\maketitle

\begin{abstract}
\textit{Background:}
Contemporary software development organizations lack diversity and the ratios of women in Free and open-source software (FOSS) communities are even lower than the industry average. Although the results of recent studies hint the existence of biases against women, it is unclear to what extent such biases influence the outcomes of software development tasks. 

\textit{Objective:}
This study aims to conceptually replicate two recent studies investigating gender biases in FOSS communities  \textit{ to identify whether the outcomes of or participation in code reviews (or pull requests) are influenced by the gender of a developer.}
In particular, this study focuses on two outcome aspects (i.e., code acceptance, and review interval) and one participation aspect (i.e., code review participation) of code review processes. 

\textit{Method:}
We will augment the dataset used in the original studies with code reviews /pull requests created during recent years. Using this dataset,  we will train multivariate regression models to accurately model the influences of developers' genders on code acceptance, review intervals, and code review participation.

\end{abstract}


\section{Introduction}
\label{sec:intro}
According to US labor market census in 2019, women make up approximately 18.8\% of technical roles in the software industry and 18\% of computer science graduates\cite{labor_market}. 
However, the number of females contributing to Free and Open Source Software (FOSS) projects is even lower (i.e., less than 10\%)~\cite{terrell_1,vasilescu2015gender}; which indicates a significant gender imbalance within the FOSS communities.  Due to the lack of gender diversity many female developers hesitate to join FOSS projects, as one woman quotes, \textit{“I feel like it’s a circle I can’t get into. But mostly I fear the excessive spotlight of being a sole female programmer on a publicly available project. In light of how women are treated on the internet, this fear does not seem unreasonable"}~\cite{open_source_women}. 
The lack of diversity among FOSS communities has been subjected to several prior studies~\cite{oss_floss, floss_pols,DAVID2008364,bosu-esem-2019}.  To increase diversity through recruitment of the underrepresented groups, FOSS communities such as Mozilla~\cite{mozilla-diversity}, Debian~\cite{debian-diversity}, and Fedora~\cite{fedora-diversity} as well as commercial software development organizations such as Google\footnote{\url{https://diversity.google/}} and Deloitte~\cite{div_deloitte}  have adopted several initiatives. 

However, these diversity initiatives may not achieve adequate result if a community does not promote `inclusion' by providing equal access to opportunities and resources for people who might otherwise be excluded or marginalized~\cite{div_inclusion}. Vernā Myers, a leading diversity and inclusion strategist, says, \textit{“While diversity means being invited to the party, inclusion means to be asked for a dance”}\cite{verna_}. Sense of inclusion drives people to be more productive and engaged with workplace as well as improves overall team performance and creativity~\cite{verna_,div_inclusion}. Organizations that practice inclusion generate up to 30\% higher revenue per employee and greater profitability than their competitors \cite{div_deloitte}. 

Despite of  these known benefits, increasing diversity and inclusion among FOSS communities are  difficult due to various explicit~\cite{floss-wiki} and implicit biases~\cite{Beatrice18} against the minorities. For example, a 2017 study by Terrell \textit{et} al.~\cite{terrell_1} found  that a woman's pull-requests  are less likely to be accepted than those created by men, if her gender is visibly identifiable. On the other hand, if a woman's gender is not identifiable, she is more likely to have a higher pull request acceptance rate than men.
Another study by Bosu \textit{et} al.~\cite{bosu-esem-2019} investigated the state of diversity and inclusion among ten popular FOSS projects. Their results hint possible biases against woman developers in three out of the ten projects as women had lower code acceptance rates as well as longer delays in getting feedback for submitted changes. Additionally in one project women received  significantly lower invitations by their peers to participate in code reviews. 

However, based on the results of these two studies~\cite{bosu-esem-2019,terrell_1}, we cannot predict to what extent developers' genders influence those outcomes,  since several other confounding factors may have been influential as well. For example, the acceptance of a pull request may be influenced by its size, complexity, and whether it includes a new feature or a bug fix. Bosu \textit{et} al.'s study~\cite{bosu-esem-2019} did not account for any of these confounding factors. Although, Terrell \textit{et} al.~\cite{terrell_1} conducted post hoc analysis to identify the influences of other factors such as  experience, patch size, patch type, and status within the community, it is unclear how much variances in acceptance rates are influenced by the genders of the contributors. To answer these questions, this study aims to conduct  a partial conceptual replication~\cite{shull2008role} of these two studies (i.e. Terrell \textit{et} al.~\cite{terrell_1}, and Bosu \textit{et} al.~\cite{bosu-esem-2019}) by  adopting measures to control for various confounding factors. 
In our proposed  replications, we aim \textit{ to identify whether the outcomes of or participation in code reviews (or  pull requests) are influenced by the gender of a developer.}

With this goal, we select two outcome aspects and one participation aspect of code reviews: i) acceptance rate-- the ratios of submitted code reviews (or pull requests) that are accepted,  2) review interval -- the time from the beginning to the end of the review process, and  3) code review participation-- the number of code reviews within a period as a reviewer. We will develop three types of regression models, where each of those three aspects would be the dependent variables and in addition to the genders of the contributors, various confounding factors would be independent variables (Table~\ref{table:variable_descriptions}). We will use \textit{multivariate regression modeling} techniques, since those can be used to estimate the relationships between a dependent variable and one or more independent variables (aka `predictors') and  are commonly used to  eliminate effects of confounding variables~\cite{pourhoseingholi2012control,harrell2015regression}.  If our study can demonstrate that code review processes in FOSS communities are significantly influenced by the genders of the contributors, it will allow project managers and owners to take extra measures to eradicate such biases. Moreover, as we plan to augment the dataset used in the original studies with code reviews/pull requests  created during the recent years (i.e., we plan to include all reviews pull requests created before May 31st, 2021), this replication will also help us to identify if awareness created through those studies has helped mitigating biases.



\section{Original Studies}
\label{sec:original-studies}

We will conduct partial conceptual replication of the studies conducted by Bosu \textit{et} al \cite{bosu-esem-2019} and Terrell \textit{et} al \cite{terrell_1}. In this replication, we aim to retest only the results suggesting explicit gender biases. The following subsections, briefly describe those two studies.

\subsection{Gender bias on Github by Terrell et al.~\cite{terrell_1}}
The main research question explored by Terrell \textit{et} al is:

\textbf{RQ:} To what extent does gender bias exist among people who judge GitHub pull requests?

They gathered the GHTorrent dataset and resolved the gender of the contributors using GitHub accounts and Google+ profiles. They found out female developer's pull requests are more likely to be accepted both as insiders (developers who are explicitly authorized as project owners and collaborators) and outsiders (GitHub users). They also examined male and female developers' pull request acceptance rates over time and showed female developers have higher acceptance rates as they become more experienced. To evaluate if women's pull request fulfill any immediate need, they examined descriptions of pull requests and the percentage of requests that refers to any issue. Results showed that women contribute comparatively less regarding immediate project needs. The authors also investigated the amount of women's contribution using three metrics: lines of code changed, number of modified files, and number of commits in each pull request. They found out that women make larger pull requests than men. Furthermore, they showed that female developers possess a higher acceptance rate in programming languages. Interestingly, they found that female developers have a lower acceptance rate when they are outsiders and their gender is identifiable. 

\subsection{Diversity and inclusion study by Bosu et al.~\cite{bosu-esem-2019}}
\textbf{RQ1:} What are the percentages of female contributors in popular OSS projects?

\textbf{RQ2:} What percentage of leadership roles are occupied by
females among the OSS projects?

\textbf{RQ3:} Are female developers of OSS projects significantly
more/less productive than their male colleagues?

\textbf{RQ4:} Is there any explicit bias for or against female
developers in open-source projects?

To answer these questions, researchers mined the code review history of 10 popular OSS projects listed in Table~\ref{table:demographic-project} and resolved genders of the active contributors by genderComputer, Google+, Facebook accounts, and Google search. They found that in all of the projects, less than 10\% of the non-casual (submitted at least five pull requests) developers are female, and also on average 4.14\% of the core developers are female. It shows a poor representation of females in leadership roles of FOSS projects. To find out if female developers are less productive, authors considered three metrics: code changes per month, code churn per month, and code review participation, and showed that there is no significant difference between male and female groups regarding contribution through code. They also found that male developers participate more in code reviews. As code review is done based on invitations, authors assume that female developers may not be invited frequently for code review. While investing bias against female developers, authors found that in three of ten projects, female developers have a lower code acceptance rate, they have to wait significantly longer to get their initial feedback from the reviewers and also they have to wait a significantly longer period of time to get their complete review. On the contrary, in three other projects, the result shows bias against male developers.

\section{Hypotheses}
\label{research_questions}
 Our main goal is to investigate whether the outcomes of or participation in code reviews are influenced by the gender of a developer. 
  Although Terrell \textit{et} al~\cite{terrell_1} found higher acceptance rates for women on Github,  Bosu \textit{et} al.'s ~\cite{bosu-esem-2019} results support that finding in only two out of the ten projects and three projects from their study suggested the contrary. Therefore, our first research question is:
 \begin{boxedtext}
\textbf{RQ1:} \emph{
 Do the genders of the contributors influence acceptance of their code changes?}
\end{boxedtext}

While we acknowledge that there are several possible choices for the gender of a person\footnote{\url{https://www.healthline.com/health/different-genders}}, due to the lack of a reliable automated mechanism to identify genders that do not fall into binary choices (e.g. man, woman), prior Software Engineering (SE) studies~\cite{vasilescu2015gender,terrell_1,bosu-esem-2019,lin2016recognizing} confined to comparisons between men and women while studying the impact of gender. Therefore, we plan to limit our investigations to identify the differences between men and women by formalizing this research question into the following two hypotheses:

\boldmath{$H1.1_0$:} \textit{There are no significant differences in code acceptance rates between men and women.}

\boldmath{$H1.1_a$:} \textit{There are significant differences in code acceptance rates between men and women.}

Since  Terrell \textit{et} al. ~\cite{terrell_1} found that women have a significantly lower code acceptance rate than men when their gender is identifiable, we pose two additional hypotheses.

\boldmath{$H1.2_0$:} \textit{There are no significant differences in code acceptance rates between men and women when gender is identifiable.}

\boldmath{$H1.2_a$:} \textit{There are significant differences in code acceptance rates between men and women when gender is identifiable.}

 \vspace{4pt}

`Code review interval' spans from the beginning to the end of the code review process~\cite{def_interval}. Bosu \textit{et} al. ~\cite{bosu-esem-2019} found that women developers had to wait three times longer than men in two out of the ten projects. However, two of their projects suggested the opposite. Our next research question aims to investigate those differences.

\begin{boxedtext}
\textbf{RQ2:} \emph{
 Do the genders of the contributors influence code review intervals for their code changes?}
\end{boxedtext}

We formalize this research question into the following two hypotheses:

\boldmath{$H2.1_0$:} \textit{There are no significant differences in review intervals between men and women.}

\boldmath{$H2.1_a$:} \textit{There are significant differences in code review intervals between men and women.}

Since identifiable gender may influence code review intervals, we pose two additional hypotheses.

\boldmath{$H2.2_0$:} \textit{There are no significant differences in code review intervals between men and women when gender is identifiable.}

\boldmath{$H2.2_a$:} \textit{There are significant differences in code review intervals between men and women when gender is identifiable.}

\vspace{4pt}
Although women did not exhibit lower productivity than men, Bosu \textit{et} al. ~\cite{bosu-esem-2019} found significantly lower participation of women in code reviews in one project. They hypothesize that, since participation in a code review is based on invitations from peers, a man may prefer other men over woman contributors for his reviewing code changes~\cite{bosu-esem-2019}. Hence, our next research question is:

\begin{boxedtext}
\textbf{RQ3:} \emph{
 Do the genders of the contributors influence their participation in code reviews?}
\end{boxedtext}

We formalize this research question into the following two hypotheses:

\boldmath{$H3.1_0$:} \textit{There are no significant differences in code review participation between men and women.}

\boldmath{$H3.1_a$:} \textit{There are significant differences in code review participation between men and women.}

However, as identifiable gender may influence code review invitations, we pose the following two hypotheses.

\boldmath{$H3.2_0$:} \textit{There are no significant differences in code review participation between men and women when gender is identifiable.}

\boldmath{$H3.2_a$:} \textit{There are significant differences in code review participation between men and women when gender is identifiable.}

\section{Execution plan}
\label{sec:execution-plan}

The following subsections detail our replication protocol, dataset collection, dependent and independent variables, analysis procedures, and potential threats to validity. 

\subsection{Replication protocol}
Replications are crucial in the software engineering domain to build knowledge through a family of experiments~\cite{santos}. According to Shull \textit{et} al.~\cite{shull2008role},  SE replications falls into two categories:
\begin{enumerate}
    \item  \textit{Exact replication:} In an exact replication,  study procedures are closely followed. Exact replications can be further divided into two subcategories: i) \textit{dependent} and ii) \textit{independent}. In a \textit{dependent replication},  all the variables and conditions are kept as close to the original studies as possible. However, some of the aspects of the original study can be modified in an \textit{independent replication}  to fit a new context.
    \item  \textit{Conceptual replication:}, This category of replications uses different research methodologies to study the same set of questions.
\end{enumerate}
 
Our replications for this study fall into the `partial conceptual replication' category, since-- i)  we plan to use a different type of statistical modeling technique, ii) we will measure and account for additional confounding variables not considered in the original studies, iii) we will use updated versions of the datasets to include recent code reviews to reflect contemporary state of the projects, and iv) we are interested in partial replications to retest only the results from the original studies hinting explicit gender biases. 

\subsection{Variables}
\label{sec:variables}

Table \ref{table:variable_descriptions} lists the three dependent variables for the three types of regression models and the independent variables grouped into two categories, i.e., contributor characteristics and changeset characteristics.  Table \ref{table:variable_descriptions}  also includes description, scale, rationale, and operationalization for each of the variables. This section presents the detail of how those variables have been measured. Most of the metrics have been used in prior studies to investigate gender bias or code review characteristics in FOSS projects \cite{pull_based,terrell_1,mcintosh}. Names of the variables inside parentheses denote them in the equations for training the regression models. While several variables can be mined directly from code review repositories, others require computations. Following subsections detail computation steps for such variables.

\begin{table*}
	\caption{Descriptions and rationale of the dependent and independent variables in our regression models. We group the independent variables into two categories: i) contributor characteristics and ii) changeset characteristics. For building the `Code review participation model', we will use only the independent variables listed in the `contributor characteristics' group.}
	\centering \label{table:variable_descriptions}
	\resizebox{\linewidth}{!}{
\begin{tabular}{p{2cm}p{4cm}lp{5cm}p{3.3cm}}
\hline
\rowcolor[HTML]{D9D9D9} 
\textbf{Name}           & \textbf{Description}                                                                                                                                     & \textbf{Scale} & \textbf{Rationale}                                                                                                                                                                                                                                                                          & \textbf{Operationalization}                                                                              \\ 
\rowcolor[HTML]{ffffff} 
\multicolumn{5}{c}{\cellcolor[HTML]{ffffff}\textbf{ Independent variables}}  \\   
\multicolumn{5}{l}{\cellcolor[HTML]{C0C0C0}\textbf{Variable group: Contributor characteristics}}  \\ 
\rowcolor[HTML]{D9D9D9} 
Gender (Gender)                  & Whether a contributor is a man or woman                                                                                                                                              & Nominal: {M, F}       & This study's objective is to identify influences of gender.                                                                                                                 & See Section~\ref{gender-comp}                                                                                          \\
\rowcolor[HTML]{EFEFEF} 
 Gender neutral profile (isGenderNeutral)  & If a contributor's profile is gender neutral                         & Nominal: {1,0}       &   Terrell \textit{et} al. found identifiable gender influencing acceptance rates~\cite{terrell_1}.                                                                                                                & See Section~\ref{profile-comp}                                                                                            \\                                                                                                                                                                                                                                                                         
\rowcolor[HTML]{D9D9D9} 
Number of total commits (totalCommit) &  Number of code commits a   developer has made to a project &  Ratio  &  Changes submitted by experienced developers who have a higher number of commits can be subjected to less scrutiny and may include a lower number of mistakes than inexperienced developers.  Experienced developers are also more likely to be invited for code reviews.                                                         &  Computed from Git commit logs.\\
\rowcolor[HTML]{EFEFEF} 
Tenure (tenure)                 & Span of time developer is contributing to the project                     & Ratio          &  Developers who are working on a project for a longer period of time are more knowledgeable of the project's design, are able to identify potential reviewers quickly than newcomers~\cite{Bosu-Carver-ESEM:2014}, and are more likely to be invited as reviewers. & Section~\ref{tenure}      
\\ 

\rowcolor[HTML]{D9D9D9} 
Review experience (revExp) & The number of pull requests that a developer has examined in the project as a reviewer                                                            & Ratio          & Experienced reviewers are more likely to get invited for more reviews                                                                                                                &   Computed using SQL query on code review dataset.   
\\ \\

\multicolumn{5}{l}{\cellcolor[HTML]{C0C0C0}\textbf{ Variable group: Changeset characteristics}}  \\ 

\rowcolor[HTML]{EFEFEF} 
Patch size (patchSize)              & Number of lines modified/  added/ deleted in a code review request                                           & Ratio          & Larger changes are more likely to be buggy,  require a longer time to review~\cite{Bosu-et-al-FSE:2014} and less like to be accepted~\cite{jiang2013will}.                                                                                                                       & Mined from code review repository.                                                                                            \\
\rowcolor[HTML]{D9D9D9} 
Cyclomatic complexity (cyCmplx)   & McCabe's cyclomatic complexity                                                                                & Ratio          & When a patch set is difficult to comprehend, it will need a longer review interval and is more likely to be rejected.                                                                                                           &      We will use the  { \tt Radon} python library to compute the cyclomatic complexity of files under review.                                                                                                  \\
\rowcolor[HTML]{EFEFEF} 
Number of patchset (numPatch)      & Number of revisions executed in the submitted request & Ratio          &  Higher number of patchsets would increase review intervals as well as the probability of rejections                                                                                                                                                                                                                                                                                            &    Mined from code review repository.                                                                                                    \\
\rowcolor[HTML]{D9D9D9} 
Is bug fix (isBugFix)              & Whether a change includes a new feature or bug fix  & Nominal: {1,0}       & Bug fixes may be in an urgent need and get reviewed quickly.    & See Section~\ref{bug-fix}           \\      
\rowcolor[HTML]{EFEFEF} 
Number of files (fileCount)         & Number of files under review in review requests                                                                & Ratio          & Changes with the higher number of files involved are more likely to be defect-prone\cite{review_2} and require longer review time.    & Mined from code review repository.   \\                      
\rowcolor[HTML]{D9D9D9} 
Number of directories (dirCount)    & Number of directories in a review request where files have been modified                                   & Ratio          & Changes spread across a large number of directories are more likely to be buggy, difficult to comprehend, and require longer review time~\cite{barnett2015helping}.                      & Mined from code review repository.                        \\
\rowcolor[HTML]{EFEFEF} 
Comment volume (cmtVolume)          & Ratio of add/modified lines that are comments                                                   & Ratio          &         Well commented changes are easy to comprehend and require shorter review intervals.                                                                                                                                                                                                                                                                                  & We will use the {\tt Radon} python library to compute the number of comment lines.                         \\
\rowcolor[HTML]{D9D9D9} 
 Ratio of new files (ratioNew)             &  Ratio of newly created files to the total number of files in a patchset                                                     & Ratio        & A new file may have more issues than a file which has been under review before. As a result, a new file can have a longer review interval and a lower acceptance rate.                                                                        &                  SQL query on data mined from code review repository. \\

                     \\

\rowcolor[HTML]{C0C0C0} 
\multicolumn{5}{c}{\cellcolor[HTML]{ffffff}\textbf{Dependent Variables}}                                                                                                                                                                                                                                                                                                                                                                                                                                                                                                                                    \\
\rowcolor[HTML]{D9D9D9} 
Acceptance (isAccepted)          & Whether a code change was accepted or rejected                                                             & Nominal: {1,0}          &  Indicates the outcome of a code review or pull request.                                                                             & Mined from code review repository                \\
\rowcolor[HTML]{EFEFEF} 
Code review interval (reviewInteval)    & Time from the beginning of the code review to the end                                                             & Ratio    & Indicates whether a contributor's code gets preferential treatment in a FOSS community              &  We will compute code review interval based on the duration from when a code review request was created to when it was marked as `Merged' or `Abandonded' or 'Rejected'.                                                                                                                                                                                                                                                                                                                                                                               \\
\rowcolor[HTML]{D9D9D9} 
Code review participation (reviewPerMonth)   & Average number of code reviews where the developer participated in the project per month                                                       & Ratio          &    Indicates whether a contributor is valued by his/her peers.  &   We will compute the average code review participation per month by dividing the total number of reviews by his/her tenure.                                                                                                                                                                                                                                                                                                                                                                    \\ \hline
\end{tabular}

}

	\vspace{-8pt}
\end{table*}

\begin{figure}
    \centering
    \includegraphics[width=\linewidth]{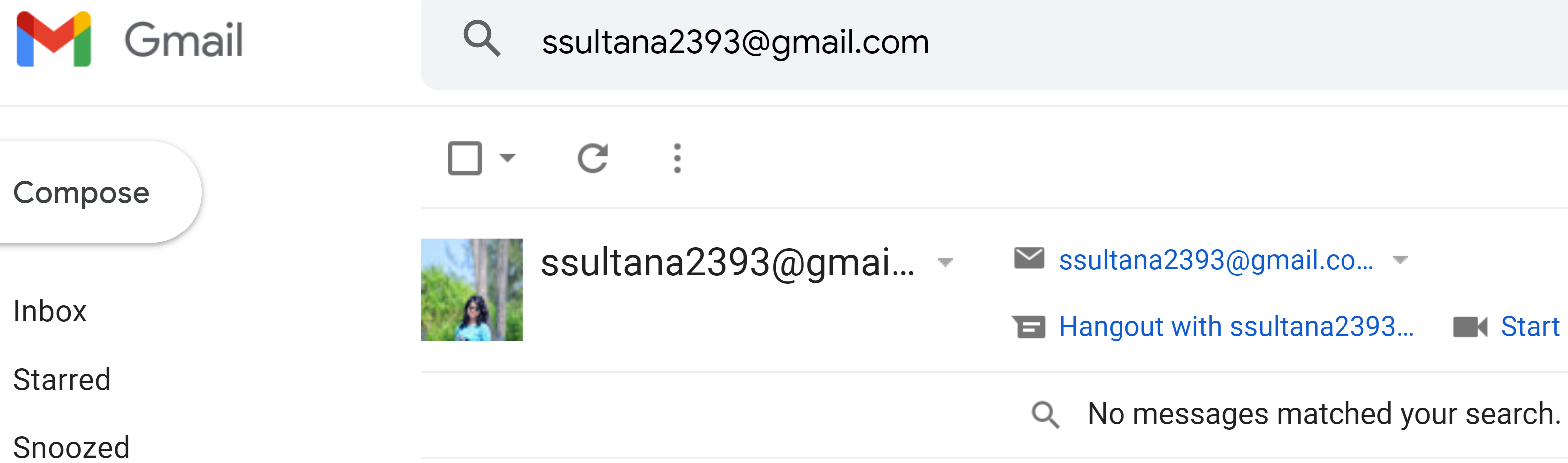}
    \caption{An example of profile search using Gmail's contact search by email address feature}
    \label{fig:contact-search}
\end{figure}

\subsubsection{Gender (Gender)}
\label{gender-comp}
Our semi-automated gender resolution steps are adopted from the methodologies used by  Terrell \textit{et} al.~\cite{terrell_1} and Bosu \textit{et} al.~\cite{bosu-esem-2019}.
Similar to those studies, we will use the genderComputer tool\footnote{\url{https://github.com/tue-mdse/genderComputer}} created by Vascilesu  \textit{et} al.~\cite{gender_v} to resolve developers' genders. The genderComputer tool classifies  a person into one of the following four categories  ( i.e., 1) Male, 2) Female, 3) Unisex, 4) None) according to that person's name. Similar to Bosu et al.~\cite{bosu-esem-2019}, we will verify the genders of the developers not marked as males, using their profiles on code review repositories (i.e., Gerrit or Github) and social networks (i.e., LinkedIn, and Facebook), However, we had to exclude the "Google Plus" search step used in the prior studies, due to the unavailability of that service. Instead of "Google plus", we will use the contact search feature on Gmail (Figure~\ref{fig:contact-search} to identify the  Google profile associated with an email address (if any).

\subsubsection{Gender-neutral profiles (isGenderNeutral):}
\label{profile-comp}
We will use the following three criteria heuristics used by Terrell \textit{et} al.\cite{terrell_1} to identify whether a profile is gender neutral.
\begin{enumerate}
    \item an auto-generated identicon is used for the profile image, as identified by ImageMagick\footnote{\url{https://imagemagick.org/index.php}}.
 \item the gender inference tool classifies the user’s both login name and display name as `unknown'.

 \item manual verification by multiple human raters fails to identify the user’s gender.
\end{enumerate}

\subsubsection{Developer's experience (totalCommit)}
\label{d-exp}
For each code review /pull request, we will count code commit experience as the number of prior code commits for the current project. We will write a SQL script to compute this attribute.

\subsubsection{Tenure (tenure)} 
\label{tenure}
We measured the tenure of a developer from the difference between the timestamp of the first submitted code commit and the most recently submitted patch for review in that particular project. Then, the differences are divided by the total number of seconds in 30 days and the ceiling values are taken to convert the integer number of months.

\subsubsection{Is bug fix (isBugFix)}
\label{bug-fix}
We will use the heuristic-based approach proposed by Zafar \textit{et} al. to identify whether a code review/pull request is for a bug fix~\cite{zafar2019towards}.

\subsubsection{Acceptance (isAccepted)}
\label{acceptance}
For every code review /pull request submitted in the project, there will be a decision if that request was merged into the project or not. We Will use this information and assign 1 for any request that was merged and 0 otherwise.

\subsection{Dataset}
\label{sec:dataset}
We will conduct our study on a total of 1010 projects, where 1,000 projects will be mined from Github based on a stratified sampling strategy. As one  of the authors of this study participated in Bosu \textit{et} al.'s~\cite{bosu-esem-2019} previous research,  we have  full access to their original dataset of 683,865 code reviews mined from 10 popular FOSS projects. We will use that dataset in our study. Projects of this dataset were selected based on two criteria: 1) Projects are actively using Gerrit. 2) Contributors of that project reviewed at least 15,000 request. Table~\ref{table:demographic-project} provides an overview of the dataset A. Although, this dataset  does not resolve several independent variables (e.g. cyclomatic complexity, is bug fix, number of files, and comment volume)  required for this study, it contains necessary information to calculate those. We will compute  those variables using the methods described in Section~\ref{sec:variables}.

\begin{table}
	\caption{Demographics of projects}
	\centering \label{table:demographic-project}
	\resizebox{\linewidth}{!}{

\begin{tabular}{|l|l|r|r|}
\hline
\textbf{Project}     & \textbf{Sponsor}           & \textbf{Request mined} & \textbf{\# of developers*} \\ \hline
Android     & Google            & 81,137         & 968                                                                \\ \hline
Chromium OS & Google            & 153,523        & 1007                                                               \\ \hline
Couchbase   & Couchbase Inc.    & 64,799         & 165                                                                \\ \hline
Go          & Google            & 19,318         & 220                                                                \\ \hline
LibreOffice & Foundation        & 35,545         & 344                                                                \\ \hline
OmapZoom    & Texas Instruments & 35,973         & 425                                                                \\ \hline
oVirt       & Redhat Inc        & 73,523         & 220                                                                \\ \hline
Qt          & Qt Company        & 155,936        & 736                                                                \\ \hline
Typo3       & Foundation        & 48,940         & 318                                                                \\ \hline
Whamcloud   & Intel             & 15,971         & 140                                                                \\ \hline

\multicolumn{4}{l}{* Number of developers, who have made at least five commits to this project.}
\end{tabular}
}

	\vspace{-19pt}
\end{table}

We will also collect code review data from another 1000 projects from Github. While the dataset used in Terrell et al.'s\cite{terrell_1} study is not publicly available, their tools and scripts are included as supplementary materials. Using those scripts, we will follow their protocols to develop subset B. However, as Github currently includes more than 65 million users\footnote{\url{https://tinyurl.com/rmcpm4ph}}, it is infeasible to include all Github users in this study. Moreover, many projects do not practice pull request based development~\cite{kalliamvakou2016depth}, which is a pre-requisite for this study. Therefore, we plan to randomly select 1000 projects that satisfies all of following four criteria.

\begin{enumerate}
   \item A software development project.
    \item Source code is available under a Open Source license.
    \item At least 10 contributors.
    \item Actively follows pull-based development with at least 20 pull requests during the last three months in our dataset.
\end{enumerate}

We will use the latest GHTorrent data dump from June 2019~\cite{Gousi13} to identify the candidate projects. We will use the Github REST API~\footnote{https://docs.github.com/en/rest} to mine latest data for the selected projects.

\subsection{Analysis Plans}

 We will use \textit{multivariate regression modeling} techniques to develop our models.
 Linear regression and logistic regression are commonly used to inspect the relation among one dependent variable and one or more independent variables when the dependent variable is scalar and binary respectively~\cite{Regression}. To develop \textit{Code acceptance} models, we will use logistic regression, since the dependent variable \textit{isAccepted} is binary.  To analyze \textit{Code Review Interval} and \textit{Code Review Participation}, we will develop linear regressions, since the dependent variables  (i.e., \textit{reviewInterval} and \textit{reviewPerMonth}) are scalars.  We have grouped the independent variables into two categories: i) contributor characteristics and ii) changeset characteristics. While both groups of independent variables will be used to train \textit{Code acceptance} and \textit{Code Review Interval} models, we will use only the in the `contributor characteristics' group to train \textit{Code Review Participation} models, since changeset characteristics are specific to a particular code change, and are not related to all the changes reviewed by a contributor during a period. Table~\ref{table:variable_descriptions} describes the variables for our models.

Recent SE studies~\cite{mcintosh,bosu_2} inspired us to adopt the models' construction and analysis approach proposed by Harrell Jr. to validate our proposed hypotheses~\cite{harrell2015regression}.
Harrell's approach will enable us to formulate non-linear relationships among variables while being aware of the possibility of overfitting (i.e., where the model performs very well for the training dataset but poorly otherwise)\cite{harrell2015regression}. We will implement Harrell's regression techniques in R using the \textit{rms} package ~\cite{R_harrell}. We detail our nine-step model development and analysis approach in the following.

\subsubsection* {Step 1: Correlation \& redundancy analysis} We will conduct correlation analysis among the independent variables of the models to remove all correlated variables. We execute the Spearman rank correlation test (\(\rho \)) that is resilient to datasets that are not normally distributed. We will also remove redundant variables constructing a hierarchical overview of the correlated variables. Among the variables residing in the same sub-hierarchy and having 
\(\mid\)\(\rho\) \(\mid\) $<$ 0.7, one will be selected and included in the final regression model. Prior studies in software engineering also considered 0.7 as the threshold value for identifying redundant variables \cite{mcintosh} \cite{Thongtanunam} \cite{bosu_2}.

\subsubsection* {Step 2: Normality adjustment} We will check if our response or dependent variable is normally distributed or not. 
For example, results of a prior study~\cite{Bosu-Carver-ESEM:2014}  suggest that distributions of code review intervals and code churns are not normally distributed. We will apply a log transformation \textit{ln(x)} to such variables\cite{mcintosh}.

 \subsubsection* {Step 3: Estimate total degree of freedom} If more degrees of freedom or independent variables are used than the dataset can afford, regression models may overfit. An overfitted model fails to show the actual relationship between dependent and independent variables. So, we calculate the budget for total degrees of freedom for our code review dataset. Harrell suggested n/15 as total degrees of freedom where n is the number of rows in dataset~\cite{harrell_2,harrell_3}.

\subsubsection* {Step 4: Allocation of degrees of freedom} We will allocate degrees of freedom to the rest of the independent variables. We intend to allocate more degrees of freedom to those variables that have higher correlations with the dependent variable.

\subsubsection* {Step 5: Equations for regression models:} Equation (1), (2), and (3) denote regression models for \textit{Code acceptance}, \textit{Code review Interval}, and \textit{Code review participation} respectively.

\vspace{-6pt}
\begin{equation}
  \begin{aligned}[b]
  & isAccepted \thicksim ln(totalCommit) + ln(patchSize) \\
  & + ln(revExp) + tenure + cyCmplx \\
  & + numPatch + isBugFix + dirCount \\
  & + cmtVolume +  ratioNew + fileCount \\
  & + gender +  isGenderNeutral
  \end{aligned}
\end{equation}
\begin{equation}
  \begin{aligned}[b]
  & ln(reviewInterval) \thicksim ln(totalCommit) + tenure\\ 
  & + ln(revExp)+  ln(patchSize)  + cyCmplx \\
  & + numPatch +isBugFix + dirCount   \\
  & + ratioNew + fileCount + gender \\
  & + isGenderNeutral + cmtVolume
  \end{aligned}
\end{equation}
\begin{equation}
  \begin{aligned}[b]
  & reviewPerMonth \thicksim tenure +isGenderNeutral \\
  & + ln(totalCommit) + gender + ln(revExp)
  \end{aligned}
\end{equation}

\subsubsection* {Step 6: Assessment of model performance} We will calculate adjusted $R^2$ to examine how well our models fit the datasets. For the \textit{Acceptance Status Model,} we will also use the Area Under the Receiver Operating Characteristic (AUC) curve to assess the performance of the model\cite{Hanley1982TheMA}.

\subsubsection* {Step 7: Estimate the power of independent variables} We will conduct Wald Chi-Square statistics (Wald $\chi^2$) to estimate the impact of each independent variable. For a particular variable, the higher the value of Wald $\chi^2$ is, the larger the impact of that variable has on the performance of the model.

\subsubsection* {Step 9: Examination of independent variables to the outcome} For linear regression models, we will plot the change of dependent variables against the change of one independent variable at a time while keeping other independent variables fixed at their median values. This plot may project how the change of any independent variable impacts the developer's code review interval or participation in reviewing. For the logistic regression model, we will calculate Odds Ratio (OR) using a 95\% confidence interval. Any independent variable with OR $>$ 1 will imply that the probability of acceptance of code (TRUE outcome) will increase with the increment of that variable.

\subsection{Validity threats}
We will examine our hypotheses for a total of 1010  FOSS projects, where 1000 projects are using Pull-request-based code reviews and the remaining 10 are using Gerrit-based code reviews. The scenario for communication and collaboration among developers may not be the same for other projects in FOSS communities. Culture differs based on the nature and size of the projects, the number of contributors, the owner of the projects, etc. So, our result might not apply to all of those projects.

\bibliographystyle{IEEEtranS}  
\bibliography{bibliography} 
\end{document}